\documentclass[reprint,
superscriptaddress,
 amsmath,amssymb,
 aps,
prb,
]{revtex4-2}

\usepackage[colorlinks,urlcolor=blue,linkcolor=blue,citecolor=blue]{hyperref}
\usepackage[table,dvipsnames]{xcolor}
\usepackage{colortbl}
\usepackage[hyphenbreaks]{breakurl}
\usepackage{color,array}
\usepackage{graphicx}
\usepackage{CJKutf8}
\usepackage{tabulary}

\usepackage{graphicx}
\usepackage{dcolumn}
\usepackage{bm}

\usepackage[utf8]{inputenc}
\usepackage[T1]{fontenc}
\usepackage{mathptmx}
\usepackage{etoolbox}
\usepackage{xcolor}
\usepackage{adjustbox}
\usepackage{float}

\begin{document}


\title{Magnetic tunnel junction as a real-time entropy source:  Field-Programmable Gate Array based random bit generation without post-processing} 

\author{Troy Criss}
\affiliation{Department of Physics, New York University, New York, New York 10003, USA}
\author{Ahmed Sidi El Valli}
\affiliation{Department of Physics, New York University, New York, New York 10003, USA}
\author{Naomi Li}
\affiliation{Department of Physics, New York University, New York, New York 10003, USA}
\author{Andrew Haas}
\email{andy.haas@nyu.edu}
\affiliation{Department of Physics, New York University, New York, New York 10003, USA}
\author{Andrew D. Kent}
\email{andy.kent@nyu.edu}
\affiliation{Department of Physics, New York University, New York, New York 10003, USA}

\date{\today}

\begin{abstract}
We demonstrate a method to generate application-ready truly random bits from a magnetic tunnel junction driven by a Field-Programmable Gate Array (FPGA). We implement a real-time feedback loop that stabilizes the switching probability near 50\% and apply an XOR operation, both on the FPGA, to suppress short-term correlations, together mitigating long-term drift and bias in the bitstream. This combined approach enables NIST-compliant random bit generation at 5~Mb/s without post-processing, providing a practical hardware solution for fast and reliable true random number generation. Beyond cryptographic applications, these capabilities open opportunities for stochastic hardware accelerators, probabilistic computing, and large-scale modeling where real-time access to unbiased randomness is essential.
\end{abstract}


\maketitle 

\section{Introduction}

Probabilistic and random bits are an essential part of several common computational problems including cryptography~\cite{McInnes1991}, Monte Carlo simulations~\cite{Harrison2010}, neural networks~\cite{Hirtzlin2019}, and the modeling of stochastic processes~\cite{Daniels2020,Misra2023}. Probabilistic computational architectures are also an emerging area of interest as an application of stochastic bits~\cite{Borders2019,Camsari2019,Camsari2023,Duffee2025}. In order for probabilistic bits to be useful for computation, they must be generated quickly, and have a high degree of precision in probability of being either 0 or 1. Most hardware True Random Number Generators (TRNGs) require additional postprocessing of their bitstreams to achieve true randomness, limiting bandwidth while taking several hours of time and enormous amounts of memory. 

\begin{figure}[b]
\centering\includegraphics[width=0.8\linewidth]{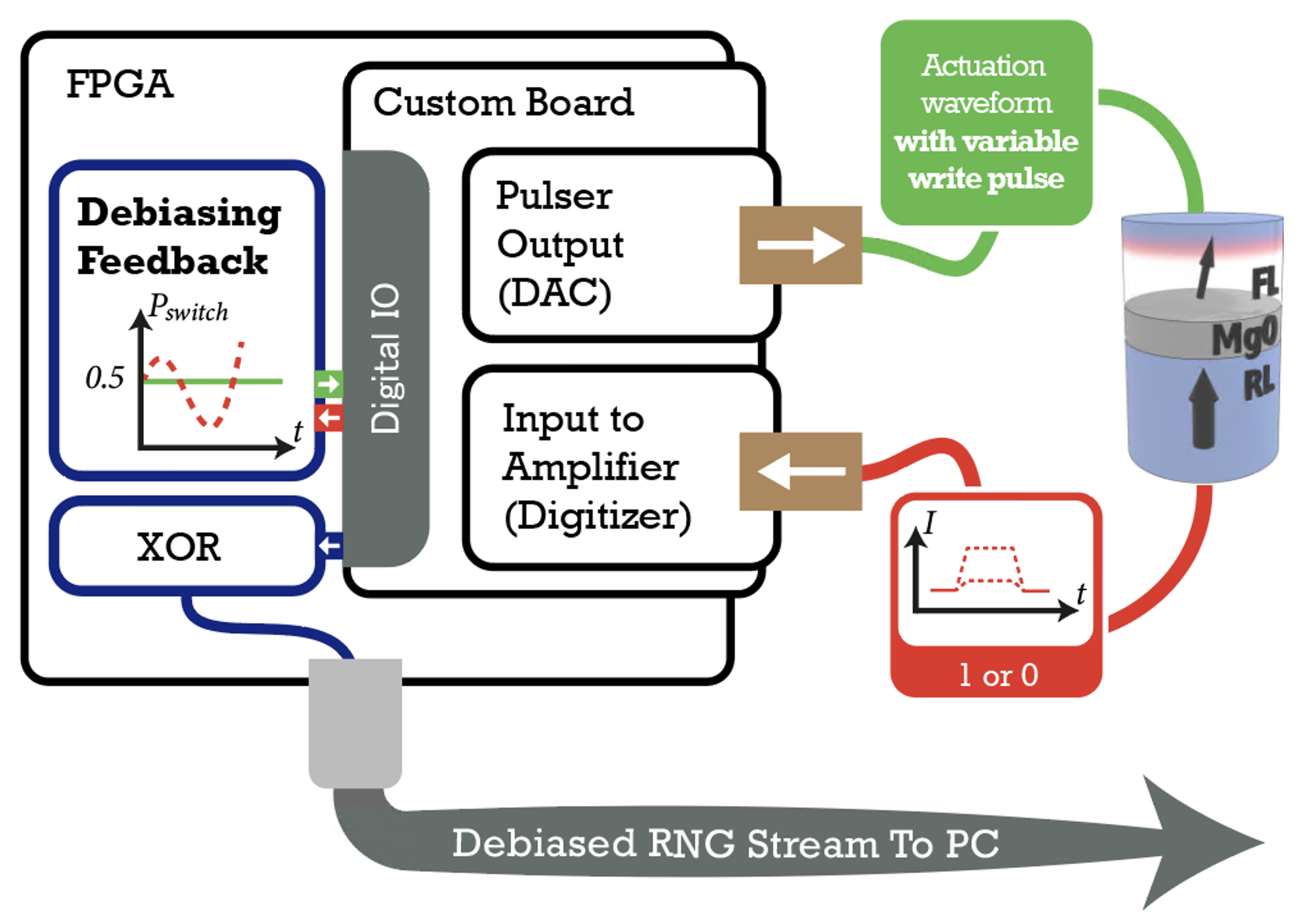}
\caption{Our custom-designed circuit uses an FPGA to control a sequence of voltage pulses that are repeatedly applied to the magnetic tunnel junction (MTJ). A transimpedance amplifier converts the small difference in current between pMTJ states into a two-level voltage signal that the FPGA registers as a stream of 0s and 1s~\cite{Dubovskiy2024}. A major improvement in this design is the inclusion of a real-time debiasing feedback loop, which continuously averages the bitstream over a defined time window and sends an Inter-Integrated Circuit (I²C) signal to the DAC to dynamically adjust the write-pulse voltage. In addition, an XOR operation is performed on the FPGA to de-correlate bits before the processed stream is transmitted to the host computer via Ethernet.}
\label{Fig:Fig1}
\end{figure}

Magnetic Tunnel Junctions (MTJs) offer a promising way to convert thermal noise into a detectable two-level electrical signals (for an up-to-date review see Ref.~\cite{Sun2025}). Shown on the right of Fig.~\ref{Fig:Fig1}, the magnetization of one of the layers of the MTJ, the free layer, is designed such that it either be ``up'' or ``down,'' so that it behaves as a two-state system. These states are separated by an energy $E_b$, and when this energy is near the thermal energy $kT$, the magnetization of the free layer of the device will fluctuate between the two states, a phenomena known as superparamagnetism. Random bits generated using superparamagnetic MTJs have been reported by a number of groups~\cite{Parks2018,Hayakawa2021,Soumah2025}. Instead of allowing for these fluctuations to occur spontaneously, stable perpendicularly magnetized MTJs (pMTJs) actuated by stochastic write pulses are another means of generating true random numbers~\cite{Rehm2023}. The device resistance after each stochastic pulse, low or high resistance, is then converted into a string of 0 and 1s. 

Previously, we have shown that we can generate and read bits at a rate of 10 Mb/s using a Field-Programmable Gate Array (FPGA)~\cite{Dubovskiy2024}. TRNGs that use pMTJs typically have correlations on short, microsecond, timescales, as well as drifts in probability over large, minute to hour timescales~\cite{Sidi2025}. These effects cause the unprocessed bit streams to fail the NIST Statistical Test Suite for randomness. In order to reduce bias and pass the NIST tests, multiple XOR operations are often employed after the data is collected. The resulting bit streams from our previous setup were able consistently pass with two XOR operations performed in Python post data collection~\cite{Dubovskiy2024}. 

In this paper we show that by implementing a feedback loop on the FPGA to modulate the stochastic write pulse sent through the pMTJ, we are able to effectively eliminate the long timescale drifts. To reduce the short timescale correlations, we introduce an XOR operation executed by the FPGA board. By changing the data acquisition conditions to enable or disable the feedback and XOR functions, we collect four distinct bitstreams from a single pMTJ. When the real-time feedback and XOR are combined, the resulting bitstream passes all NIST tests for statistical randomness, demonstrating that we have eliminated the need for post processing, and can generate application ready bits at a rate of 5 Mb/s.

\section{Experiment and Setup}

The following experiments are all conducted on the same 50 nm-diameter pMTJ, with a fixed magnetic electrode layer and a free layer that switches between ``up'' and ``down.'' More detailed descriptions of the device can be found in Ref.~\cite{Rehm2023}.
We used a custom built daughter board and FPGA dev kit (Intel Cyclone 10 GX FPGA, qDK-DEV-10CX220-A) to operate the experiment, as illustrated schematically in Fig.~\ref{Fig:Fig1}. A state machine in Verilog on the FPGA toggles analog switches (part \# SN74AUC2G66DCTR) which are connected to the 16 channels of two 8-channel digital to analog converters (DACs) (part \# DAC7578SPWR). The voltages of these channels are summed and sent through the pMTJ, and output current is measured after passing through a transimpedance amplifier (part \# OPA699ID). A more detailed description of the circuit can also be found in \cite{Dubovskiy2024}. 

\begin{figure}[t]
\centering\includegraphics[width=0.8\linewidth]{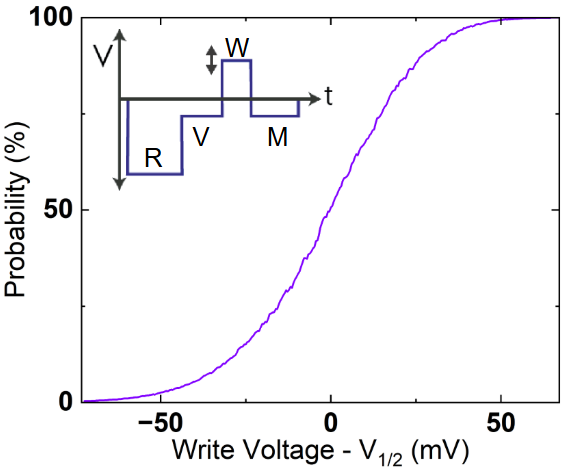}
\caption{The pulse sequence shown in the upper left inset is applied to the junction at a repetition rate of approximately 10.6~MHz to generate random bits. A large reset (R) pulse initializes the junction to a known state, which is verified by a smaller verify (V) pulse. A stochastic write (W) pulse of opposite polarity produces a random MTJ state, followed by a measure (M) pulse---identical to the verify pulse---that reads the final state. The FPGA is used to systematically increase the write-pulse amplitude in steps of 0.44~mV and the switching probability is plotted as a function of write voltage with 10 million trials per step. The voltage at which the probability of switching is 50\% is denoted as $V_{1/2}$; the x~axis shows the deviation from this voltage.}
\label{Fig:Fig2}
\end{figure}

In order to generate random bits, a write pulse is tuned so that the free layer of the pMTJ will switch states as close to 50\% of the time as possible. By sending a lower amplitude pulse, the difference in resistance between the (P) and (AP) states can be detected without disturbing the pMTJ state. The complete generation of one bit consists of the following pulse sequence: 1) a higher amplitude, positive polarity, reset pulse reliably sets the junction to a known, in this case AP, state; 2) a positive, low amplitude, verify pulse detects the state of the junction, verifying that it is in the AP state; 3) a tuned write pulse of opposite polarity puts the junction into the random state; 4) a measure pulse identical to the verify pulse reads the final state of the junction, which is recorded by the FPGA as a 0 or 1 after the signal is amplified. The characterization of a MTJ switching probability vs write voltage and a diagram of the applied pulse sequence is shown in Fig.~\ref{Fig:Fig2}.

\begin{figure}[!t]
\centering\includegraphics[width=0.8\linewidth]{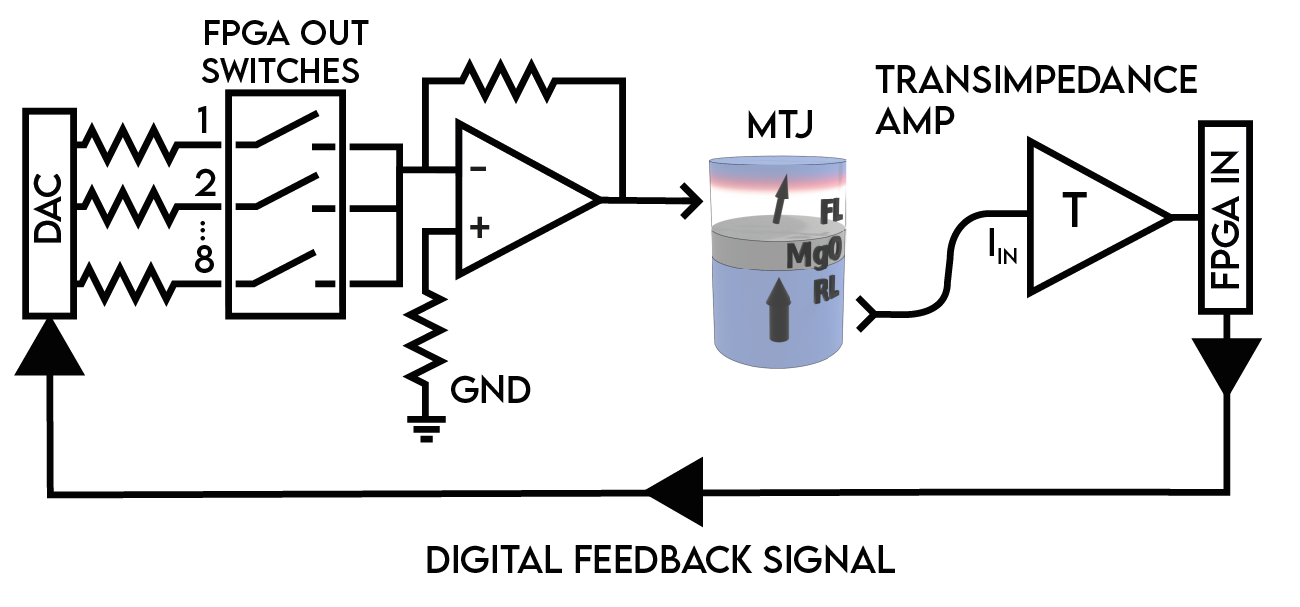}
\caption{The FPGA controls voltages on eight DAC channels, four of which are used in the pulse sequence, each gated by an analog switch. Four of the eight channels are first inverted to negative voltages before being switched, and one of these is used for the writing pulse. During the measure (M) pulse, the current through the MTJ is amplified by a transimpedance amplifier, compared to a threshold, and recorded by the FPGA as a single bit. The results from repeated trials are averaged over time, and a feedback signal adjusts the write-pulse amplitude to maintain a switching probability near 50\%.}
\label{Fig:Fig3}
\end{figure}

\subsection{Feedback Scheme}

Variation of pulse amplitudes has been shown to be an effective method of correcting for probability drift while avoiding correlations between bits and maintaining truly random behavior~\cite{Oosawa2015,Zhang2025,Koh2025}. In this section, we describe our digital FPGA implementation, which allows for easy testing and adjusting of feedback programming. 

Figure~\ref{Fig:Fig3} shows schematic of the circuit. The digital feedback signal, based on the input to the FPGA, changes the DAC output voltage. To implement feedback, we must set two parameters: the frequency at which feedback adjustments can occur, and the acceptable error within which feedback should not be applied. We need the frequency of the feedback to be much shorter than the minute to hour timescale of probability drift~\cite{Sidi2025}. However, we want to minimize correlations that changing the write pulse voltage creates. We found that below about 0.1 Hz drift can start to occur, and that above about 10 Hz correlations between bits from voltage changes begin effecting the data. We therefore chose 1 Hz for our experiments here. 

To determine the acceptable error in our bit stream, we consider the resolution of the digital-to-analog converters (DACs) that control the pulse amplitudes, and the change in probability caused by a change in write pulse voltage. We use 12-bit DACs, which can vary the amplitude of each pulse from 0 to 1.8~V, in steps of approximately 0.44~mV. With the calibration in Fig.~\ref{Fig:Fig2}, we measured that near 50\% switching, $dp/dV \approx 1.6\;\%/\text{mV}$. The minimal DAC voltage step of 0.44~mV thus results in a change in switching probability of about 0.7\%. In our experiments, we chose $\pm0.5\%$ as the acceptable error in probability, as we found that allowing greater error could cause the switching probability to settle with a small offset and remain uncorrected for multiple feedback cycles.

With these parameters, the FPGA computes the average probability bias of the bitstream once per second (\emph{i.e.}, for $10^6$ bits), and compares it to a user defined target probability. Then it sends an Inter-Integrated Circuit (I$^2$C) signal to the DACs to adjust the write voltage accordingly---increasing it by one DAC step if the switching probability is too low, and decreasing it by one step if it is too high. To generate truly random bits we set the target probability to 50\% for all of our experiments. 

\subsection{FPGA XOR}
In earlier implementations, the XOR operation used to debias the bitstream was performed offline, after the experiment had concluded. The bitstream was divided into two equal halves, and the XOR of corresponding bits from each half was computed in Python---a process that required several hours and was computationally inefficient. In the present work, the XOR operation is implemented directly on the FPGA, where it is performed continuously during the experiment \emph{before} the bits are transmitted to the PC. This modification required XOR'ing bits that are much closer together in time. However, because adjacent bits exhibit short-timescale correlations, XOR'ing bits that are too close in sequence does not effectively remove bias. In fact, XOR'ing immediately adjacent bits tended to produce probabilities greater than 50\%, as consecutive bits were more likely to be opposite. A separation of at least 4096 bits was found to be sufficient to suppress these correlations and allow the XOR operation to effectively reduce bias without excessive FPGA memory use; this bit separation was thus adopted in the experiments reported here.

\section{Results: Evaluation of Four Bitstreams}

To assess the effectiveness of the feedback loop and the on-board XOR operation, four data sets were collected using the same pMTJ device. First, a bitstream was recorded with neither feedback nor XOR enabled to establish a baseline. Second, feedback was activated without XOR to evaluate its ability to compensate for long-timescale bias drift. Third, the XOR operation was enabled without feedback to examine its effect on whitening the data. Finally, both feedback and XOR were applied simultaneously to demonstrate that the resulting bitstream can pass the NIST statistical tests when taken directly from the FPGA output. Each data set consists of $10^{11}$ bits that are averaged in  $10^5$ bits segments, which were converted to time using the bit rate: 10.6~Mb/s without XOR and 5.3~Mb/s with XOR.

\begin{figure*}[t] 
  \centering
  \includegraphics[width=\textwidth]{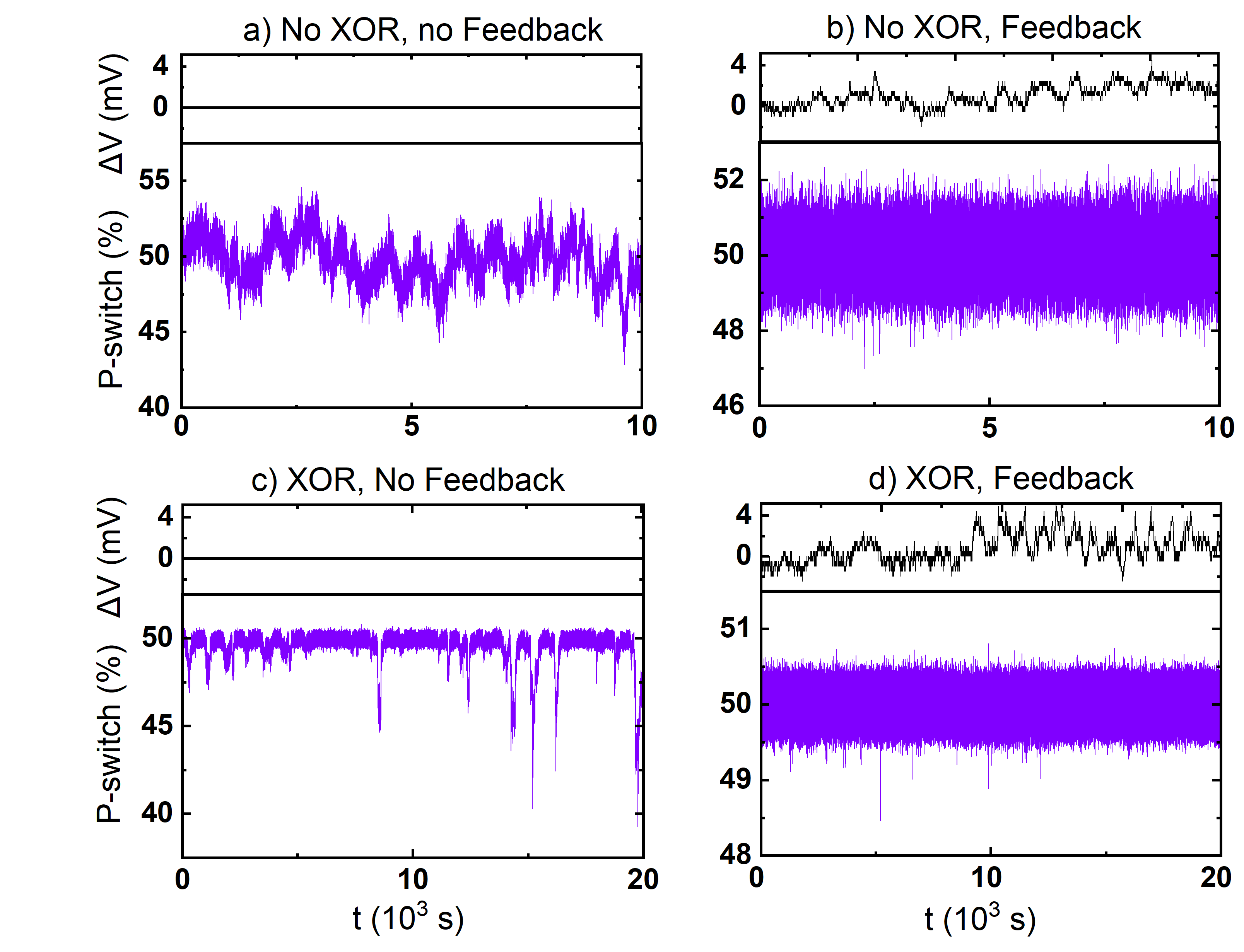}  
  \caption{Four trials varying the usage of feedback and XOR. Each trial has a total of $10^{11}$ bits, collected at a rate of 10.6 Mb/s. The results were averaged in bins of $10^5$ bits, and their index was converted to time on the x axis. The change in write pulse, $\Delta{V} = V_W - V_{1/2}$ is plotted simultaneously in the upper panel of each figure. a) No feedback no XOR: we observed large drifts in probability and large short timescale variation. b) Feedback no XOR: the long term drifts were eliminated, but the large variation remained. The write voltage increased slightly during the experiment c) No feedback XOR: Now collected at an effective rate of 5.3 MHz, the short term variation was greatly reduced, but the drift was still present. d) Feedback XOR: Both the drift and variation have been significantly reduced. Again, the write voltage increased over the course of the experiment.}
  \label{Fig:wide_plot}
\end{figure*}

\subsection{No XOR No Feedback}
In Fig.~\ref{Fig:wide_plot}(a), neither feedback nor XOR was used, and the result is consistent with previous work~\cite{Dubovskiy2024}. The constant write voltage of 358~mV is shown in the top panel, where $V_{1/2}$ denotes the voltage at the beginning of the experiment that generated a probability of 50\%, and the y-axis is the difference $\Delta V=V - V_{1/2}$. Short-timescale variations of 1--2\% arise from correlations between consecutive trials, while long-timescale drifts of 4--5\% in switching probability develop over the nearly three-hour experiment. In subsequent measurements, the XOR operation was found to suppress short-term variance, whereas the feedback loop eliminated long-term drift. When applied together, these two techniques effectively stabilized the bitstream across all timescales.

\subsection{No XOR With Feedback}

The resulting bitstream with feedback is shown in Fig.~\ref{Fig:wide_plot}(b). The switching probability remained stabilized near 50\% throughout the experiment, although short-term correlations persisted. The feedback loop effectively eliminated the 4--5\% long-timescale drifts observed in the previous measurement. To quantify this improvement, the variance of the switching probability, $\sigma^2 = \sum_i (x_i - \bar{x})^2/N_T$, where $N_T$ is the total number of bins ($N_T=10^6$). For the dataset without feedback or XOR, $\sigma^2 = 2.07$, whereas with feedback alone, $\sigma^2 = 0.319$. Thus by suppressing low-frequency ($<1$~Hz) drift in the switching probability, the feedback loop reduced the overall variance by more than an order of magnitude.

The write voltage during the experiment, shown in the top panel, increases gradually throughout the trial, reaching a value about 4~mV greater than $V_{1/2}$. From Fig.~\ref{Fig:Fig3}, the slope near 50\% switching probability is $dP/dV \approx 1.6\%\!/\text{mV}$, so a 4~mV deviation corresponds to a change in switching probability of approximately 6.5\%. This value is consistent with a slightly larger drift than observed in Fig.~\ref{Fig:wide_plot}(a), where drifts of up to about 5\% were recorded. The gradual increase in write voltage suggests that changes in the junction’s static resistance during the measurement require a higher pulse amplitude to maintain a 50\% switching probability~\cite{Sidi2025}.

\subsection{XOR No Feedback}

Our raw data has an error $ P = 0.5 +\epsilon$, where P is the target probability 0.5. In our data without feedback we observe that $\epsilon$ ranges from approximately 0.01 - 0.05. The XOR operation reduces this to $P_{XOR} = 0.5 -2\epsilon^2$~\cite{Matsui}. Because the XOR returns a 0 when given two 1's or two 0's, the probability of uncorrelated bits should always be below fifty percent, which can be seen in Fig.~\ref{Fig:wide_plot}c). The bits at the beginning of this experiment, shortly after the write voltage had been manually set to give 50 percent switching have a very low probability bias, greatly reducing the problem of short term correlations. As drift causes the bias of the raw data to increase, variations appearing as downward spikes increase in amplitude and frequency.

\subsection{XOR with Feedback}
Our previous trials showed that feedback can eliminate long period drifts, while an XOR is effective at reducing short term errors. Finally, we combined these two techniques to produce an extremely precise and stable bit stream in Fig.~\ref{Fig:wide_plot}d). Again looking at the variations of the data in Figs.~\ref{Fig:wide_plot}c) and ~\ref{Fig:wide_plot}d) we found that with XOR and no feedback $\sigma^2 = 1.01$, which is greater than the data with no XOR but with feedback, showing the importance and ensuring consistency of the bit stream. With both feedback and XOR $\sigma^2 = 0.051$, the smallest of all four trials. By combining feedback and an XOR operation on the FPGA we can thus generate bits with a precise, stable, probability at an effective rate of 5Mb/s after a single XOR. 

The write voltage vs time plot is consistent with the observations we made from Figs.~\ref{Fig:wide_plot}a) and b) with $V - V_{1/2}$ peaking at about 4~mV, again exhibiting the trend of increasing write voltage as the junction's composition changes during the experiment. 

\begin{figure}[t]
\centering\includegraphics[width=0.5\textwidth]{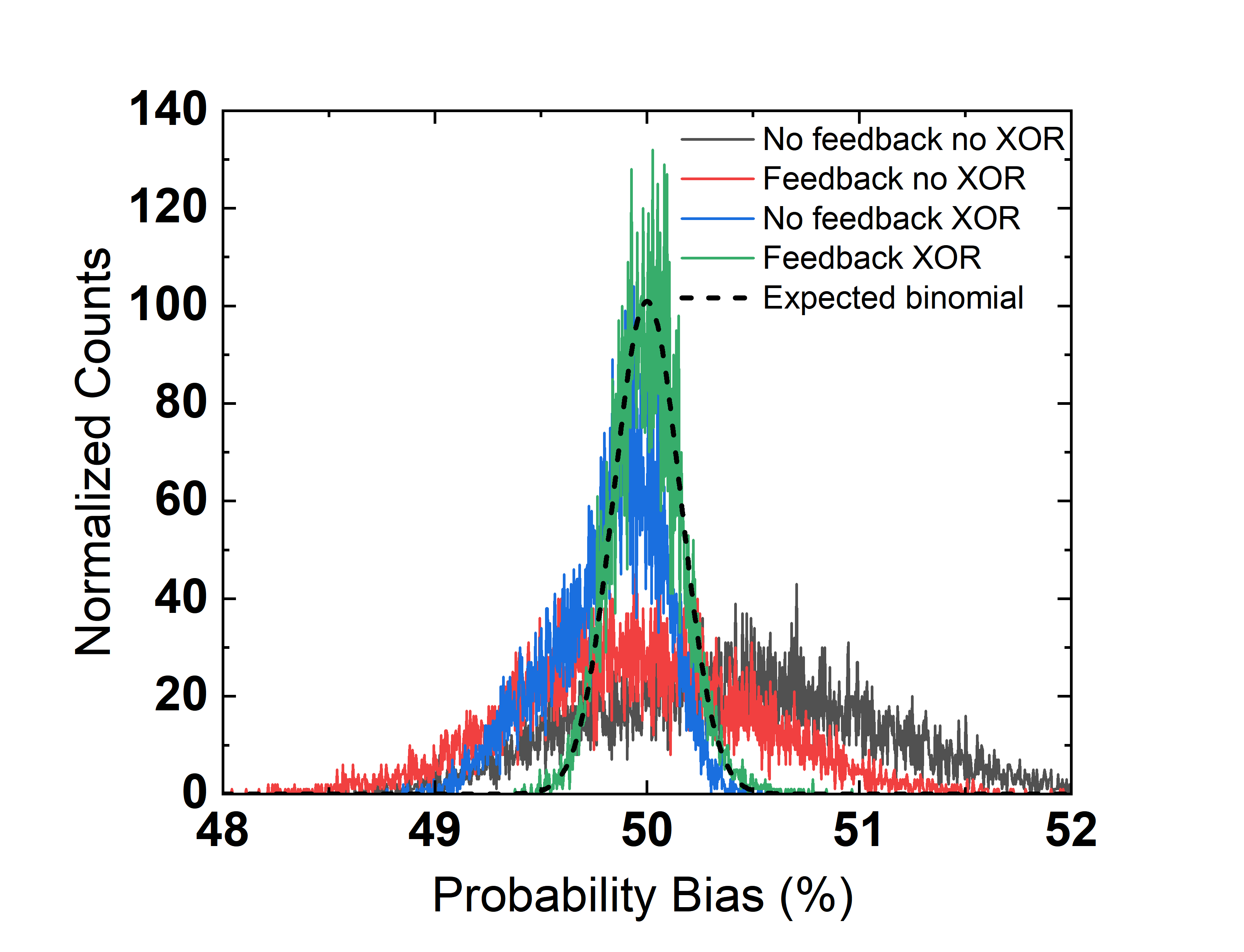}
\caption{A sample of the first billion bits from each data set are averaged in bins of $N = 10^5$ bits and placed in bins $4\times10^{-3}$ percent wide. The results are compared to the expected binomial function described in Eq. ~\ref{binomial}. Here $50$~\% is the target probability and $x\in [0,100]$. We observed that the dataset with both XOR and feedback best fits the expected distribution.} 
\label{Fig:Hist}
\end{figure}

\section{Analysis}

While we have demonstrated the ability of combining feedback to reduce the variance and error of our data, further analysis is required to determine if the data meets criteria for randomness. We begin by plotting the probability density of each experiment. A sample of the first $10^9$ bits are averaged in bins of $N=10^5$ bits and placed in $T=10^4$ bins, $\mathrm{d}x=4\times10^{-3}$ percent wide. We then generated a normal probability density function, defined by the bin size $N$ and the target probability $p=50$~\% and normalized it to the area and range of each distribution. The density function is
\begin{equation}
    f(x) \mathrm{d}x= \frac{T}{\sqrt{2\pi\sigma^2}} \exp{\left(-\frac{(x-\mu)^2}{2\sigma^2}\right)\mathrm{d}x},
    \label{binomial}
\end{equation}
where $\sigma^2 = p(100-p)/N$ and $\mu = 50$ and $x \in [0,100]$. 

In Fig.~\ref{Fig:Hist} the histogram counts are plotted for each dataset and compared to the expected function (Eq.~\ref{binomial}). The datasets without an XOR have a much greater variance than the expected distribution, though the shift right in the raw data is corrected by the feedback no XOR data, which is centered much closer to 50 percent. We also observe the effect of the XOR skewing the data to the left in the no feedback XOR histogram. However, with both feedback and XOR we are able to generate a distribution that closely matches the expected normal probability density function.

\begin{figure*}[t]
  \centering
  \includegraphics[width=\textwidth]{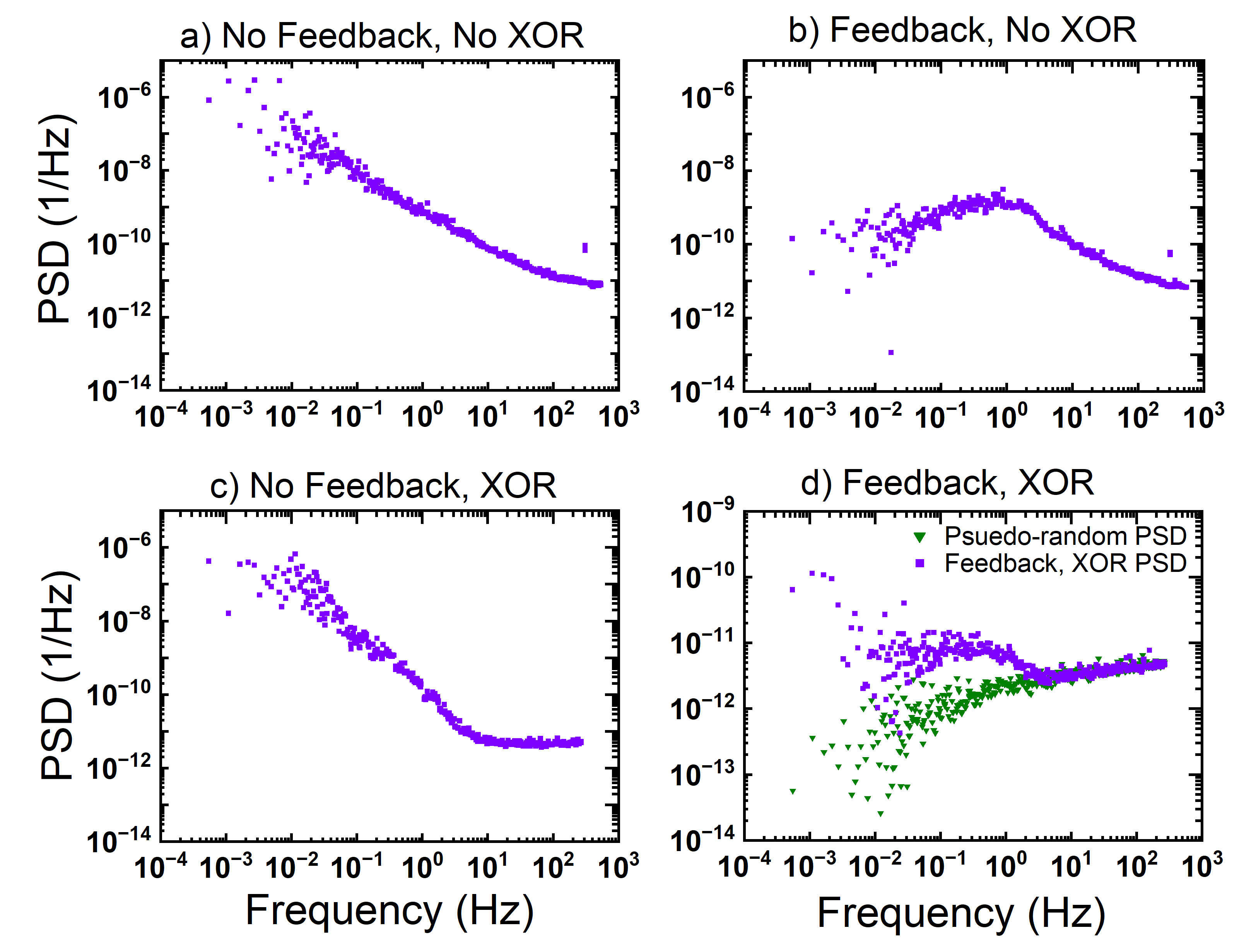} 
  \caption{Power spectral density (PSD) of probability for each run. In a) the raw data has a power spectrum consistent with thermal noise. In b) the implementation of feedback reduces the PSD below the feedback frequency of 1~Hz .Using just an XOR operation in figure c) increases the slope of the power spectrum, and generates white noise above about 10~Hz. Combining the feedback that whitens the noise at low frequencies and the XOR that whitens the noise at high frequencies we get the closest spectrum to that of white noise in d). To demonstrate the power spectrum's consistency with random noise we compare with the PSD of bits generated by Python's random.getrandbits() function. Note that the vertical scale is expanded in d).}
  \label{fig:FFT}
\end{figure*}

Next, we perform an FFT on the each dataset to observe the amplitude of probability fluctuations at different frequencies. To capture the low frequency fluctuations the bits were averaged in bins of 10,000 bits and the power spectral density (PSD) of the average probability of these bins was taken. The envelope of the PSD spectrum is plotted in Fig.~\ref{fig:FFT} in the frequency $10^{-4}$ to $10^3$ Hz 

In Fig.~\ref{fig:FFT}a) we can see that the FFT of the data with no feedback and no XOR exibits a Brownian noise spectrum. In Fig.~\ref{fig:FFT}b) the spectrum is smoothed closer to white noise below the feedback frequency of 1~Hz, as random fluctuations below that frequency are corrected. In both cases without an XOR there is a small peak at just over 100 Hz, likely coming from an electrical component in the circuit, which we are still investigating. 

When the XOR is turned on without feedback the spectrum flattens from brown noise to white noise at a lower frequency than before, at around 5 Hz, displayed in Fig.~\ref{fig:FFT}c). Finally in Fig.~\ref{fig:FFT}d), plotting on a narrower scaled, the XOR whitens the high frequencies and the feedback whitens the low frequencies to achieve a PSD most similar to white noise. The PSD of pseudo-random bits generated using Python’s standard random.getrandbits() function (which employs the Mersenne Twister algorithm) is overlaid for comparison. At medium to high frequencies, our measured PSD closely matches that of the pseudo-random sequence, indicating comparable statistical flatness in this frequency range, while preserving the true physical randomness inherent to our device-based bit generation.

Lastly, to determine if our data is usable for applications calling for probabilistic bits, we use the NIST statistical test suite for randomness~\cite{Bassham2010}. To run the tests, a sample of 1 million bits were taken from near the halfway point of each trail, in Table~\ref{Nist} the passing tests are marked in green while the failing tests are marked red, and the sum of the total passed tests is tallied in the bottom row. A pass is defined as passing the given test on 95 percent or more of the trials. The experiments without XOR consistently perform poorly on the same tests, and score similarly overall. When the XOR is introduced the data performs much better in almost all tests, though the long timescale fluctuations that are present by the halfway point of our experiment cause some tests to continue to fail. Adding the feedback along with the XOR results in a bit stream that passes all tests throughout the entire duration of the experiment, achieving our goal of truly random bit generation using just one XOR operation conducted on the FPGA. 

\begin{table}[h]
\caption{NIST Statistical Test Suite results on all four datasets.}
\footnotesize
\centering
\renewcommand{\arraystretch}{1.2} 
\setlength{\extrarowheight}{0pt}
\begin{tabular}{ |>{\centering\arraybackslash}m{2.5cm}||
                 >{\centering\arraybackslash}m{1.3cm}|
                 >{\centering\arraybackslash}m{1.3cm}|
                 >{\centering\arraybackslash}m{1.3cm}|
                 >{\centering\arraybackslash}m{1.3cm}|}
\hline
\textbf{Test name} & No feedback no XOR & Feedback no XOR & No feedback XOR & Feedback XOR \\
\hline
Frequency & 
\cellcolor{red!50}18/100 & \cellcolor{red!50}21/100 & \cellcolor{red!50}40/100 & \cellcolor{green!50}96/100 \\
Block Frequency&
\cellcolor{red!50}91/100 & \cellcolor{red!50}90/100 & \cellcolor{green!50}98/100 & \cellcolor{green!50}99/100 \\
Runs & \cellcolor{red!50}31/100 & \cellcolor{red!50}28/100 & \cellcolor{red!50}80/100 & \cellcolor{green!50}98/100 \\
Longest Run
& \cellcolor{green!50}99/100 & \cellcolor{green!50}97/100 & \cellcolor{green!50}100/100 & \cellcolor{green!50}100/100 \\
Rank & 
\cellcolor{green!50}98/100 & \cellcolor{green!50}100/100 & \cellcolor{green!50}100/100 & \cellcolor{green!50}97/100 \\
FFT & 
\cellcolor{red!50}90/100 & \cellcolor{red!50}90/100 & \cellcolor{green!50}99/100 & \cellcolor{green!50}97/100 \\
Non-Overlapping &
\cellcolor{red!50}109/148 & \cellcolor{red!50}95/148 & \cellcolor{green!50}148/148 & \cellcolor{green!50}148/148 \\
Overlapping & \cellcolor{red!50}67/100 & \cellcolor{red!50}83/100 & \cellcolor{green!50}100/100 & \cellcolor{green!50}100/100 \\
Maurer's & 
\cellcolor{green!50}99/100 & \cellcolor{green!50}100/100 & \cellcolor{green!50}99/100 & \cellcolor{green!50}99/100 \\
Linear Complexity & \cellcolor{green!50}100/100 & \cellcolor{green!50}99/100 & \cellcolor{green!50}99/100 & \cellcolor{green!50}98/100 \\
Serial & \cellcolor{green!50}195/200 & \cellcolor{green!50}197/200 & \cellcolor{green!50}198/200 & \cellcolor{green!50}197/200 \\
Approximate Entropy & \cellcolor{red!50}65/100 & \cellcolor{red!50}63/100 & \cellcolor{green!50}100/100 & \cellcolor{green!50}99/100 \\
Cumulative Sums & \cellcolor{red!50}10/200 & \cellcolor{red!50}36/200 & \cellcolor{red!50}86/200 & \cellcolor{green!50}192/200 \\
\hline
Totals &
1072/1548\cellcolor{red!50} & 
1099/1548\cellcolor{red!50} & 
1347/1548\cellcolor{red!50} & 
1520/1548\cellcolor{green!50} \\
\hline

\end{tabular}
\label{Nist}
\end{table}

\section{Conclusion} 

We have demonstrated with our updated setup including a real time feedback system and XOR implemented on an FPGA can generate truly random bits at a rate of 5 Mb/s. We have eliminated the need for post processing and created a device that can stream ready to use bits directly on to a host computer for applications such as cryptography, Monte Carlo simulations, and stochastic modeling. To further increase the bit rate, new pulse schemes that employ two stochastic write pulses without a reset or verify pulse can increase the data rates~\cite{Fu2023}. Also, the pulse durations can be decreased by at least an order of magnitude. For example, in a different setup with a Teledyne FPGA without feedback or XOR functions, we have demonstrated 100~Mb/s data rates~\cite{Sidi2025}. We are also interested in alternative debiasing algorithms to the XOR, which may retain more bits while providing a smaller debiasing factor. Further processing can also be implemented on the FPGA, for example the NIST tests can be done on the FPGA to continuously check that the data is truly random. Our setup provides an adaptable platform to further test and optimize true random number generation with MTJs, as well as to explore other applications such as generating probability distributions by dynamically changing the target probability, and generating extremely low probability bias bitstreams to simulate rare events. 

\section*{Data Availability}
\noindent
The data that support the findings of this study are available from the corresponding authors upon reasonable request.

\begin{acknowledgments}
We acknowledge support from the Office of Naval Research (ONR) under Award
No. N00014-23-1-2771. We thank Andre Dubovski for helpful discussions of the setup and approach.
\end{acknowledgments}

\bibliography{smtj_lib}

\end{document}